\documentstyle[12pt,epsfig]{article}
\begin{document}
\font\ninerm = cmr9

\def\footnoterule{\kern-3pt \hrule width \hsize \kern2.5pt}

\def\ru1{\rule[-0.4truecm]{0mm}{1truecm}}
\newcommand{\gapproxeq}{\lower .7ex\hbox{$\;\stackrel{\textstyle >}{\sim}\;$}}
\newcommand{\lapproxeq}{\lower .7ex\hbox{$\;\stackrel{\textstyle <}{\sim}\;$}}

\pagestyle{empty}

\begin{flushright}
CERN-TH/2000-034\\
hep-ph/0001305 \\
$~$ \\
January 2000
\end{flushright}

\vskip 0.5 cm

\begin{center}
{\Large\bf Testing $\kappa$-Poincar\'e 
with neutral kaons}
\end{center}
\vskip 1.5 cm
\begin{center}
{\bf Giovanni AMELINO-CAMELIA\footnote{{\it Marie Curie
Fellow} of the European Union
(address from February 2000: Dipartimento di Fisica,
Universit\'a di Roma ``La Sapienza'',
Piazzale Moro 2, Roma, Italy)}}
and {\bf Franco BUCCELLA\footnote{On leave of absence from
Dipartimento di Scienze Fisiche,
Universit\'a di Napoli, Mostra D'Oltremare, Pad.~19,
Napoli, Italy.}}\\
\end{center}
\begin{center}
{\it Theory Division, CERN, CH-1211, Geneva, Switzerland}
\end{center}

\vspace{1cm}
\begin{center}
{\bf ABSTRACT}
\end{center}

{\leftskip=0.6in \rightskip=0.6in

In recent work on experimental tests of
quantum-gravity-motivated phenomenological models,
a significant role has been played by the so-called ``$\kappa$''
deformations of Poincar\'e symmetries.
Sensitivity to values of the
relevant deformation length $\lambda$
as small as $5 \cdot 10^{-33}m$ has been achieved in recent
analyses comparing the structure of $\kappa$-Poincar\'e 
symmetries with data on the gamma rays we detect
from distant astrophysical sources.
We investigate violations of CPT symmetry which may be associated
with $\kappa$-Poincar\'e in the physics of the neutral-kaon system.
A simple estimate indicates that experiments on the neutral kaons
may actually be more $\lambda$-sensitive than
corresponding astrophysical experiments,
and may already allow to probe values of $\lambda$ of order
the Planck length.

}

\newpage
\baselineskip 12pt plus .5pt minus .5pt
\pagenumbering{arabic}
\pagestyle{plain} 


It has been recently realized~\cite{gacgrb,schaef,billetal}
that observations of gamma rays from distant astrophysical sources
can be used to set stringent bounds on the length parameter that
characterizes ``$\kappa$-Poincar\'e'', one of the most
studied scenarios
for the dimensionful quantum deformation of Poincar\'e symmetries,
which has been developed most notably
in Refs.~\cite{LNRT:def,shahnkappamin,kpoin}.
This realization has attracted significant attention because
the level of sensitivity of planned gamma-ray observatories
(see, {\it e.g.}, Refs.~\cite{AMS,GLAST})
should be sufficient to explore values of
the length parameter as small as the Planck length
($L_p \sim 1.6 \cdot 10^{-35}m$).
Only very few other experimental
contexts~\cite{ehns,koste,gacgwi,polonpap}
can achieve this type of ``Planckian sensitivity'',
and such sensitivity levels with respect to $\kappa$-Poincar\'e
are especially meaningful in light of the fact that
it appears likely~\cite{koste,thooft,aemn1,gacgrf98,gampul,berto}
that the space-time foam of quantum gravity would induce
a dimensionful (although, of course, not necessarily ``$\kappa$'')
deformation of Poincar\'e symmetries.

We are here interested in exploring the consequences of the fact
that~\cite{gacmaj} the $\kappa$-Poincar\'e deformation
would also induce a corresponding deformation of CPT invariance.
There is a wide literature on the
idea that ordinary CPT invariance might be violated
by quantum-gravity effects (see, {\it e.g.},
Refs.~\cite{ehns,koste,pagegaume,hpcpt,emnk,floreacpt}),
but usually the relevant quantum-gravity scenarios
are not developed to the point of allowing a definite
prediction for the effects of violation
of ordinary CPT invariance, and one is led
to the use of multi-parameter phenomenological approaches.
As observed in Ref.~\cite{gacmaj},
in the $\kappa$-Poincar\'e framework it appears plausible
that one should arrive at a definite prediction for all
CPT violating effects parametrized only by the
single length parameter $\lambda$
characterizing\footnote{Note that in some of the related 
literature the length parameter
we denote by $\lambda$ is written as $\hbar \kappa^{-1}$
(see, {\it e.g.}, Refs.~\cite{LNRT:def,shahnkappamin,kpoin})
or $\hbar c E_{QG}^{-1}$
(see, {\it e.g.}, Refs.~\cite{gacgrb,schaef,billetal}).
Following Ref.~\cite{gacmaj}, we adopt the $\lambda$ notation
to emphasize the forefront role played by the $\kappa$-Minkowski
space-time in our considerations.
In particular, a requirement of duality with respect to
$\kappa$-Minkowski singles out~\cite{shahnkappamin,kpoin}
the realization of $\kappa$-Poincar\'e
which is considered here and in 
Refs.~\cite{gacgrb,schaef,billetal,gacmaj}.}
the underlying non-commutative ``$\kappa$-Minkowski''
space-time~\cite{shahnkappamin}
\begin{equation}
[x^j,t]=\imath \lambda \, {x^j \over c} , \quad [x^j,x^k]=0
\label{mink}
\end{equation}
where $c$ is the conventional
speed-of-light constant and $j,k=1,2,3$.

Such a single-parameter description of CPT violation within
the $\kappa$-Poincar\'e framework however still requires~\cite{gacmaj}
the development of several mathematical tools on the
non-commutative $\kappa$-Minkowski space-time.
Since it appears likely that a long time will be necessary
for these technical developments to mature,
in this Letter we present a simple heuristic estimate 
of $\kappa$-Poincar\'e-induced
CPT violation in the neutral-kaon system.
As already emphasized in Ref.~\cite{gacmaj},
the neutral-kaon system, with its delicate balance
of mass/length scales, can provide a natural context
for setting bounds on $\lambda$ through
tests of CPT invariance.
It is in fact
well established~\cite{ehns,pagegaume,hpcpt,emnk,floreacpt}
that observable properties of the neutral-kaon system are extremely
sensitive to any deviation from ordinary CPT invariance.

Our heuristic estimate is based on the
$\kappa$-deformed dispersion 
relation~\cite{gacgrb,LNRT:def,shahnkappamin,kpoin,gacmaj}
\begin{equation}
c^4 M^2 = {\hbar^2 c^2 \over \lambda^{2}}
\left(e^{\lambda E \over \hbar c}
+e^{-{\lambda E \over \hbar c}}
-2 \right)- c^2 \vec{p}^2 e^{-{\lambda E \over \hbar c}}
\simeq E^2 - c^2 \vec{p}^2 
+ {\lambda c E \over 2 \hbar} \vec{p}^2 ~,
\label{disp}
\end{equation}
where $\hbar$ is the Planck constant, $M$, $\vec{p}$ and $E$
respectively denote the mass,
the momentum and the energy of the particle,
and, since available data already imply~\cite{billetal}
$\lambda < 5 \cdot 10^{-33}m$,
on the right-hand side we used the fact that
in all contexts of possible interest the particles will
satisfy $E \ll \hbar c/\lambda$.
Clearly, solving for $E$ in Eq.~(\ref{disp})
one does not recover the ordinary result (with its traditional
two solutions of equal magnitude and opposite sign); instead,
one finds that the two solutions $E_+$, $E_-$ are given by
\begin{equation}
E_{\pm} \simeq - {\lambda c \over 2 \hbar} \vec{p}^2 \pm
\sqrt{c^4 M^2+ c^2 \vec{p}^2} ~.
\label{sols}
\end{equation}
Since it is anyway quite natural for quantum gravity to violate CPT
invariance~\cite{ehns,pagegaume,hpcpt,emnk,floreacpt} and
Eq.~(\ref{sols})
provides not exactly opposite solutions for $E_+$ and $E_-$,
as an heuristic $\kappa$-Poincar\'e-inspired argument we conjecture that
in the phenomenological Hamiltonian describing neutral-kaon
dynamics in presence of CPT violation
within otherwise ordinary quantum mechanics\footnote{The possibility
of CPT violation outside quantum mechanics
has also been considered in the literature~\cite{emnk},
and it is not implausible~\cite{gacmaj} that it might turn out
to be the proper way to describe CPT violation in the
$\kappa$-Poincar\'e framework; however, for the present
preliminary and heuristic analysis we shall only
consider the possibility of CPT violation within quantum mechanics.}
\begin{equation}
  H = \left( \begin{array}{cc}
 (M + {1\over 2} \delta M) - {1\over 2} i(\Gamma + {1\over 2}
 \delta \Gamma)
&   M_{12}^{*} - {1\over 2} i\Gamma _{12}^{*} \\
           M_{12}  - {1\over 2} i\Gamma _{12}
&    (M - {1\over 2}
    \delta M)- {1\over 2} i(\Gamma
    - {1\over 2}
    \delta \Gamma ) \end{array}\right)\ ,
\label{hmatr}
\end{equation}       
the parameter $\delta M$
could take a value of order 
\begin{equation}
|\delta M| \sim {¦ E_{+}¦ - ¦ E_{-}¦ \over ¦ E_{+}¦ + ¦ E_{-}¦} 2 M
 \simeq {\lambda c \over \hbar} {\vec{p}^2 M \over 
\sqrt{c^4 M^2+ c^2 \vec{p}^2}} ~.
\label{stimadm}
\end{equation}
We also make an heuristic estimate of the parameter $\delta \Gamma$
using the observation that a difference $\delta M$
in the masses would induce a corresponding difference for the rate
into two pions, which gives the main contribution to the width
of the opposite-strangeness kaons.
The amplitude is proportional to $M^2 - M_\pi^2$
(it should vanish in the SU(3) limit~\cite{cabi,gellmann})
\begin{equation}
\Gamma \sim
{\sqrt{M^2 -  4 M_\pi^2} (M^2 - M_\pi^2)^2
\over M^2} ~
\label{stimag}
\end{equation}
and (if $\delta \Gamma$ is exclusively due to $\delta M$)
this leads to the estimate 
\begin{equation}
{|\delta \Gamma| \over \Gamma} \sim \left(
{3 M^4 - 7 M^2 M_\pi^2 - 8 M_\pi^4 
\over M^4 - 5 M^2 M_\pi^2 + 4 M_\pi^4} \right) { \delta M \over M} 
\simeq 4 { \delta M \over M} ~,
\label{prostimadg}
\end{equation}
which according to (\ref{stimadm}) corresponds to
\begin{equation}
|\delta \Gamma|  \sim  4 {\lambda c \over \hbar} 
{\vec{p}^2 \Gamma \over 
\sqrt{c^4 M^2+ c^2 \vec{p}^2}} ~.
\label{stimadg}
\end{equation}

Based on the structure of our estimate (\ref{prostimadg}),
the fact that $\Gamma/M \ll 1$
and the fact that in precision measurements on the neutral-kaon
system $\delta M$ competes
with the quantity $M_{K_L} - M_{K_S} \sim 3.5 \cdot 10^{-15} GeV$
and $\delta \Gamma$ competes 
with $\Gamma \sim 7.4 \cdot 10^{-15} GeV$
we conclude that the bounds on $\lambda$ that can be derived
from our estimate (\ref{stimadg}) of $\delta \Gamma$
are necessarily much less stringent than the corresponding
bounds obtainable from our estimate (\ref{stimadm}) of $\delta M$.
We restrict our attention to (\ref{stimadm})
in the following.

A prominent feature of our estimate (\ref{stimadm})
is that, for given $|\lambda|$, it predicts a $|\delta M|$
which is an increasing function of $|\vec{p}|$,
quadratic in the non-relativistic limit
and linear in the ultra-relativistic one.
Therefore among experiments achieving comparable
$\delta M$ sensitivity the ones studying more energetic kaons
are going to lead to more stringent bounds on $\lambda$.

Meson factories in $e^+ e^-$ rings, such
as DA$\Phi$NE~\cite{maianicpt,cptdafne},
can provide strong tests of CPT~\cite{koste}.
In particular, the KLOE experiment is expected
to reach~\cite{cptdafne}
sensitivity to values of $\delta M$ around $3 \cdot 10^{-18} GeV$,
which, using our estimate (\ref{stimadm}) and the fact that
the kaons have momenta of about $110 MeV$
at the $\phi$ resonance, corresponds to sensitivity
to values of $|\lambda|$ around $6 \cdot 10^{-32}m$.

Of course, the argument for $\kappa$-Poincar\'e-induced
CPT violation
we have considered for the neutral-kaon system would also apply
to the other analogous neutral-meson systems
and, despite the lower precision reachable for $\delta M_B$
at beauty factories~\cite{babar,jb}, 
sensitivity comparable to the one of KLOE
could be achieved as a result of the boost with respect to
the laboratory, which is a consequence of the asymmetric
setup of these factories.
At BaBar~\cite{babar}
electrons with momenta $\sim \! 9 GeV$ colliding with positrons
with momenta $\sim \! 3 GeV$ produce B's with momenta $\sim \! 2.8 GeV$
and this, for the expected sensitivity~\cite{coco} 
to $\delta M_B \sim 5 \cdot 10^{-15} GeV$,
provides sensitivity to values
of $|\lambda|$ around $1.4 \cdot 10^{-31}m$.

A better sensitivity
is given by the CPLEAR experiment~\cite{cplear},
which studied the neutral kaons produced in the reactions 
\begin{equation}
p {\bar p} \rightarrow K^+ \pi^- {\bar K}^0 ~,~~~~
p {\bar p} \rightarrow K^- \pi^+ K^0 ~,
\label{reactions}
\end{equation}
with the strangeness of the neutral kaon being tagged as the opposite
of the corresponding charged kaon.
They find~\cite{cplear}
$|\delta M| < 7.5 \cdot 10^{-18} GeV$
and this corresponds to the
bound $|\lambda| < 1.2 \cdot 10^{-32} m$, 
taking into account that
the CPLEAR reactions happen at rest so that the
neutral-kaon energy is 
around $2 M_{proton}/3$, which corresponds to a momentum
of order 0.4 GeV.
By neglecting $\delta \Gamma$ and considering only the contribution
of the $\pi \pi$ decay channel to the
Bell-Steinberger unitarity relationship~\cite{bsref}
\begin{equation}
\left[ {1 \over 2} (\Gamma_{K_S} + \Gamma_{K_L}) 
+ i (M_{K_L} - M_{K_S}) \right] <K_S|K_L> = \Sigma_f 
A^*(K_S \rightarrow f)
A(K_L \rightarrow f)
\label{bsrel}
\end{equation}
CPLEAR obtains~\cite{cplear} the bound 
$|\delta M| < 4.4 \cdot 10^{-19} GeV$
and this corresponds to the
bound $|\lambda| < 7 \cdot 10^{-34} m$.

Very stringent bounds on $\lambda$
can be derived from the data
of the NA31~\cite{na31} and the E731~\cite{fermilab}
experiments
by assuming the approximate equality
\begin{equation}
M_{{\bar K}^0} - M_{K^0} \simeq 2 {(M_{K_L} - M_{K_S}) |\eta|
\left( {2\over 3} \phi^{+-} + {1\over 3} \phi^{00} - \phi_0 \right)
\over sin \phi_0}
\label{unita}
\end{equation}
where the phases $\phi^{+-}$ and $\phi^{00}$
are defined by
\begin{equation}
{A(K_L \rightarrow \pi^{+(0)} \pi^{-(0)}) 
\over A(K_S \rightarrow \pi^{+(0)} \pi^{-(0)})} = |\eta^{+-(00)}|
e^{i \phi^{+-(00)}}
~,
\label{fidef}
\end{equation}
and $\phi_0$ is the superweak phase
\begin{equation}
\phi_0 = tg^{-1} {2 (M_{K_L} - M_{K_S}) \over 
\Gamma_{K_S} - \Gamma_{K_L} } = 43.5^o \pm 0.1^o
~,
\label{bsdef}
\end{equation}
which also follows from unitarity
by considering only 
the $\pi \pi$ decay channel on the right-hand side of 
Eq.~(\ref{bsrel}) and neglecting any other source of CPT violation.
The experimental results~\cite{na31}
\begin{equation}
\phi^{+-} = 46.9 \pm 2.2^o~,~~~~~
\phi^{00}=47.1 \pm 2.8^o
~,
\label{nares}
\end{equation}
and~\cite{fermilab}
\begin{equation}
\phi^{+-} = 42.4 \pm 1.4^o~,~~~~~
\phi^{00}-\phi^{+-}=-1.6 \pm 1.2^o
~,
\label{eres}
\end{equation}
lead to the bounds 
$|\delta M| < 5 \cdot 10^{-18} GeV$
and
$|\delta M| < 2.5 \cdot 10^{-18} GeV$
respectively.
Taking into account that the kaon beam 
has an average momentum of $100 GeV/c$ in NA31
and ranges from $40 GeV/c$ to $150 GeV/c$ in E731
(which we take to correspond to a reference momentum
of $80 GeV/c$)
one obtains the bounds $|\lambda| < 2 \cdot 10^{-35} m$
from NA31 data and 
\begin{equation}
|\lambda| < 1.2 \cdot 10^{-35} m
~
\label{newbound}
\end{equation}
from E731.\footnote{The fact that the sample 
of the NA31 and E731 data
is rather homogenous (both measuring the phases $\phi^{+-}$ 
and $\phi^{00}$ and using kaons with momenta of roughly the
same order) can provide motivation for obtaining a limit
on $\lambda$ by combining the results of the two experiments.
By properly taking into account the slightly different
kaon-momentum scales of the two experiments we find 
the combined limit $|\lambda| < 7 \cdot 10^{-36} m$.
The fact that the central values of the NA31 and of the E731
data would correspond to values of $\lambda$ with opposite signs
is one of the factors that contribute to 
rendering the combined limit significantly 
(roughly a factor 2) more stringent than (\ref{newbound}).}
This ``conditional'' bound on $\lambda$ (conditional in the sense
that it relies on an heuristic analysis which might turn out to be
unreliable when tested within a more rigorous study)
obtained by analyzing the neutral-kaon system
goes even beyond the Planck length
and is significantly more
stringent than the previous bound $\lambda < 5 \cdot 10^{-33}m$,
which was obtained
by comparing~\cite{gacgrb,schaef,billetal}
data on the gamma rays we detect from distant astrophysical
sources with the structure of the $\kappa$-Poincar\'e
deformed dispersion relation.
Our analysis therefore provides motivation
for more rigorous analyses of the implications
of $\kappa$-Poincar\'e for the neutral-kaon system; in fact,
if our heuristic estimate was confirmed by such more rigorous studies
one could conclude that precision measurements on the
validity of CPT in the neutral-kaon system
are a better probe of the $\kappa$-Poincar\'e deformation than
the astrophysical observations previously considered in the literature.

This possible ``competition'' between astrophysical and neutral-kaon-related
bounds on $\lambda$ is made possible by the fact that,
as already emphasized in Ref.~\cite{gacmaj},
the $\kappa$-Poincar\'e framework can lead to phenomenological models
in which the magnitude of all new effects could
be related directly and calculably to the single parameter $\lambda$,
while in other quantum-gravity-motivated formalisms
the evaluation of the magnitude of the effects
directly from the original theory turns out to be too difficult and
one can only make phenomenological
models~\cite{koste,aemn1,emnk}
to parametrize the magnitude of the effects, 
with independent parametrizations
for each of the effects.

Looking beyond our relation (\ref{stimadm}),
which is only a conjecture in the framework 
of $\kappa$-Poincar\'e, and also in light of the fact that
$\kappa$-Poincar\'e is of course only one specific
candidate for quantum-gravity-deformed symmetries,
it is worth observing that in any quantum-gravity theory
predicting a non-vanishing $\delta M/M$
and predicting an effect that is linear in a characteristic
length scale $\lambda$ (to be possibly identified with the
Planck length) 
one would be able to write a relation
of the type $\delta M /M \sim \lambda E^*/\hbar $,
where $E^*$ carries dimensions of an energy
(from now on we set $c = 1$ to simplify formulas).
Our conjecture within the $\kappa$-Poincar\'e framework 
corresponds to the case $E^* = \vec{p}^2 /E$, but in general
$E^*$ will be given by some combination of $M$, $|\vec{p}|$ and $E$.
For the simple possibilities
$E^* = E$,
$E^* = |\vec{p}|$,
$E^* = M$,
in addition to the possibility $E^* = \vec{p}^2 /E$ 
we already considered,
we report in Table 1 the limits on $\lambda$
which can be obtained (using unitarity)
from the quoted experiments.
Notice that, apart from the case $E^* = M$ (where the best bound 
is set by CPLEAR), E731 always sets the best bound 
on $\lambda$ and this
bound is more than two orders of magnitude
better than the astrophysical bound $\lambda < 5 \cdot 10^{-33}m$
(the kaons of E731 and NA31
are relativistic and therefore do 
not distinguish between the scenarios 
$E^* = E$,
$E^* = |\vec{p}|$ and
$E^* = \vec{p}^2 /E$,
as it is evidently also true for the gamma rays used 
for the astrophysical bound).
Even the bounds set by CPLEAR are  
more stringent than 
the astrophysical bound.

\begin{table}[p]
\begin{center}
\begin{small}
\begin{tabular}{||c||c|c|c|c||}
\hline
$~$ & $E^* = \vec{p}^2 /E$ & $E^* = E$ 
& $E^* = |\vec{p}|$ & $E^* = M$ \\
\hline\ru1 
DA$\Phi$NE & $6 \cdot 10^{-32}$ &  $2.8 \cdot 10^{-33}$
& $1.3 \cdot 10^{-32}$ & $2.8 \cdot 10^{-33}$ \\
\hline\ru1 
BABAR &  $1.4 \cdot 10^{-31}$    & $3.6 \cdot 10^{-32}$ 
& $7.2 \cdot 10^{-32}$ & $ 4 \cdot 10^{-32}$ \\
\hline\ru1 
CPLEAR & $7 \cdot 10^{-34} $  &  $2.8 \cdot 10^{-34}$
& $ 4.4 \cdot 10^{-34}$ & $ 3.6 \cdot 10^{-34}$ \\
\hline\ru1 
NA31 & $2 \cdot 10^{-35}$   & $2 \cdot 10^{-35}$ 
& $2 \cdot 10^{-35}$ & $4 \cdot 10^{-33}$ \\
\hline\ru1 
E731 & $1.2 \cdot 10^{-35}$   & $1.2 \cdot 10^{-35}$ 
& $1.2 \cdot 10^{-35}$ & $2 \cdot 10^{-33}$ \\
\hline
\end{tabular}
\end{small}
\end{center}
\label{table} 
\caption{Limits on $\lambda$ (in meters)
that can be obtained (using unitarity)
from DA$\Phi$NE, BABAR, CPLEAR, NA31 and E731
for the cases $E^* = \vec{p}^2 /E$,
$E^* = E$,
$E^* = |\vec{p}|$ 
and
$E^* = M$.} 
\end{table}

In closing, we observe that
the study of neutral kaons and other neutral mesons might also
prove useful for the exploration of some of the open conceptual issues
associated with deformations of spacetime symmetries,
and in particular with $\kappa$-Poincar\'e.
Most notably, above our estimates were obtained by considering
the momentum of the particles in the laboratory frame of reference
(which also made sense for the comparison with the astrophysical bounds
where a corresponding assumption was made), but the frame dependence
inherently associated with $\kappa$-Poincar\'e renders this choice
quite significant. One could for example imagine that
the ``$\kappa$-Poincar\'e-preferred frame'' would be set by the 
particular state of the quantum-gravity 
foam~\cite{gampul,wheely,hawkfoam,weave}
realized in the experiment.
If effects of the type here considered were eventually
detected it seems
likely that the controlled environment of neutral-meson experiments
might prove more useful than astrophysical experiments
for the investigation of these delicate issues.

\bigskip
\bigskip
\bigskip
\bigskip
It is our pleasure to greatfully acknowledge conversations
on these topics with Giuseppe Nardulli and Holger Nielsen.

\baselineskip 12pt plus .5pt minus .5pt


\begin{thebibliography}{99}

\bibitem{gacgrb} G. Amelino-Camelia, J. Ellis, N.E. Mavromatos, 
D.V. Nanopoulos and S. Sarkar, 
{\it Tests of quantum gravity from observations 
of $\gamma$-ray bursts}, 
astro-ph/9712103,
Nature {393} (1998) 763.

\bibitem{schaef} B.E.~Schaefer,
{\it Severe limits on variations of the speed 
of light with frequency},
Phys.~Rev~Lett.~82 (1999) 4964.

\bibitem{billetal} S.D. Biller {\it et al},
{\it Limits to Quantum Gravity Effects from Observations
of TeV Flares in Active Galaxies},
Phys.~Rev.~Lett.~83 (1999) 2108. 

\bibitem{LNRT:def}
J.~Lukierski, A.~Nowicki, H.~Ruegg, and V.N. Tolstoy,
Phys. Lett. B264 (1991) 331.

\bibitem{shahnkappamin} S.~Majid and H.~Ruegg, 
Phys.~Lett.~B334 (1994) 348.

\bibitem{kpoin} 
J.~Lukierski, A.~Nowicki and H.~Ruegg, 
Ann.~Phys.~243 (1995) 90.

\bibitem{AMS} AMS Collaboration, S.~Ahlen {\it et al.}, 
Nucl.~Instrum.~Meth.~A350 (1994) 351.

\bibitem{GLAST}  GLAST Team, E.D.~Bloom {\it et al.},
{\it Proc.~Intern.~Heidelberg Workshop on TeV Gamma-ray
Astrophysics}, eds. H.J.~Volk and F.A.~Aharonian (Kluwer, 1996)
pp.109-125.

\bibitem{ehns}
J.~Ellis, J.S.~Hagelin, D.V.~Nanopoulos and M.~Srednicki,
Nucl.~Phys.~B241 (1984) 381.

\bibitem{koste} V.A.~Kostelecky and R.~Potting,
Phys.~Rev.~D51 (1995) 3923.

\bibitem{gacgwi} G. Amelino-Camelia, 
{\it Gravity-wave interferometers as quantum-gravity detectors},
gr-qc/9808029,
Nature {398} (1999) 216.

\bibitem{polonpap} G. Amelino-Camelia, CERN-TH/99-223,
gr-qc/9910089,
{\it Are we at
the dawn of quantum-gravity phenomenology?}, notes based 
on lectures given at the XXXV Karpacz Winter School of Theoretical
Physics
{\it From Cosmology to Quantum Gravity}, Polanica, Poland, 2-12
February, 1999 (to appear in the proceedings).

\bibitem{thooft}
G.~`t Hooft,
Class.~Quant.~Grav.~13 (1996) 1023.

\bibitem{aemn1} 
G.~Amelino-Camelia, J.~Ellis, N.E.~Mavromatos and D.V.~Nanopoulos, 
Int. J. Mod. Phys. {A12} (1997) 607.

\bibitem{gacgrf98} G. Amelino-Camelia, 
Mod. Phys. Lett. {A13} (1998) 1319.

\bibitem{gampul}
R.~Gambini and J.~Pullin,
{\it Nonstandard optics from quantum space-time},
Phys.~Rev.~D59 (1999) 124021.

\bibitem{berto} O.~Bertolami and C.S.~Carvalho, gr-qc/9912117.

\bibitem{gacmaj} G.~Amelino-Camelia and S.~Majid,
hep-th/9907110.

\bibitem{pagegaume}  D.N.~Page, 
Gen.~Rel.~Grav.~14 (1982) 299;
L.~Alvarez-Gaume and C.~Gomez,
Commun.~Math.~Phys.~89 (1983) 235.

\bibitem{hpcpt}
P.~Huet and M.E.~Peskin,
Nucl.~Phys.~B434 (1995) 3.

\bibitem{emnk}
J.~Ellis, J.~Lopez, N.E.~Mavromatos and D.V.~Nanopoulos,
Phys.~Rev.~D53 (1996) 3846.

\bibitem{floreacpt} F.~Benatti and R.~Floreanini, 
Nucl.~Phys.~B488 (1997) 335.

\bibitem{cabi} N.~Cabibbo, Phys.~Rev.~Lett.~12 (1964) 62. 

\bibitem{gellmann} M.~Gell-Mann, Phys.~Rev.~Lett.~12 (1964) 155.

\bibitem{maianicpt} L.~Maiani, in {\it The Second DA$\Phi$NE
physics handbook},
L.~Maiani, G.~Pancheri and N.~Paver eds.~(INFN, Frascati, 1995).

\bibitem{cptdafne} V.S.~Demidov and E.P.~Shabalin,
in {\it The DA$\Phi$NE physics handbook},
L.~Maiani, G.~Pancheri and N.~Paver eds.~(INFN, Frascati, 1992).

\bibitem{babar} The BaBar collaboration, Technical Design Report
SLAC-R-95-457 (March 1995).

\bibitem{jb} F.~Takasaki,
{\it Status of KEKB accelerator and detector},
presented at the International Symposium on Lepton and
Photon Interactions at High Energies, Stanford (California) 1999.

\bibitem{coco} P.~Colangelo and G.~Corcella, 
Eur.~Phys.~J.~C1 (1998) 515.

\bibitem{cplear} A.~Angelopoulos et al, {\it $K_0-\bar{K_0}$ mass
and decay-width differences: CPLEAR evaluation},
report CERN-EP/99-150 (submitted to Phys.Leet.B).

\bibitem{bsref} J.S.~Bell and J.~Steinberger,
in {\it Proceedings of the Oxford Conference 
on Elementary Particle Physics} (1995) 195.

\bibitem{na31} R.~Carosi et al, 
Phys.~Lett.~B237 (1990) 303.

\bibitem{fermilab} M.~Woods et al, 
Phys.~Rev.~Lett.~60 (1988) 1695;
J.R.~Patterson et al, 
Phys.~Rev.~Lett.~64 (1990) 1491.

\bibitem{wheely} J.A.~Wheeler, {\it Relativity, groups and topology},
ed.~B.S. and C.M.~De Witt (Gordon and Breach, New York, 1963).

\bibitem{hawkfoam} S.W.~Hawking, 
Nuc.~Phys.~B144 (1978) 349.

\bibitem{weave} A.~Ashtekar, C.~Rovelli and L.~Smolin,
Phys.~Rev.~Lett.~69 (1992) 237.

\end{thebibliography}
\end{document}